 \documentclass[final,5p,times,twocolumn,num]{elsarticle}

\usepackage{amssymb}
\usepackage{amsmath}
\usepackage{lipsum}
\usepackage{xcolor}

\journal{Journal of Alloys and Compounds}

\begin{document}

\begin{frontmatter}



\title{How does picosecond structural deformation of (Ba,Sr)TiO$_{3}$ relate to the pyroelectric effect?}


\author[TXp]{D. Schmidt}
\affiliation[TXp]{organization={TXproducts UG (haftungsbeschraenkt)}, addressline={Luruper Hauptstr. 1, 22547 Hamburg}, country={Germany}}

\author[IFW,TUD]{J. Wawra}%
\affiliation[IFW]{organization={Leibniz Institute for Solid State and Materials Research}, addressline={Helmholtzstrasse 20, 01069 Dresden}, country={Germany}}
\affiliation[TUD]{organization={Institute for Applied Physics, Technische Universitaet Dresden}, addressline={01062 Dresden}, country={Germany}}

\author[IKZ]{D. Hensel}
\affiliation[IKZ]{organization={Leibniz-Institut fuer Kristallzuechtung}, addressline={Max-Born-Str. 2, 12489 Berlin}, country={Germany}}

\author[IKZ]{M. Brede}

\author[IFW]{R. H{\"u}hne}%

\author[IKZ,TXp]{P. Gaal}
\ead{peter.gaal@ikz-berlin.de}
\date{\today}

\begin{abstract}
The pyroelectric effect in ferroelectric thin films is typically composed of different contributions, which are difficult to disentangle. In addition, clamping to the substrate interface plays an important role. We studied epitaxial (Ba,Sr)TiO$_3$  thin films grown on NdScO$_3$ to see if time-resolved measurements can shed more light on the complex interaction. In particular, we compare standard measurements of the pyroelectric coefficient by temperature-dependent hysteresis loops to transient deformation measurements on picosecond timescales in the same material. The advantage of the time-resolved approach lies in its increased sensitivity in thin films compared to that of polarization hysteresis measurements. Whereas a fast thermal expansion of the ferroelectric thin film was observed after femtosecond laser excitation of the intermediate SrRuO$_3$ layer, heat diffusion simulations reveal frustration of the thermal expansion, which might be explained with the charge dynamics at the Schottky barrier formed at the SrRuO$_3$/(Ba,Sr)TiO$_3$. More studies are required to quantitatively assess the individual contributions to the pyroelectric coefficient of the materials used in our layer architecture.
\end{abstract}



\begin{keyword}
ferroelectric \sep time-resolved x-ray diffraction \sep pyroelectric effect \sep structural dynamics



\end{keyword}

\end{frontmatter}




\section{\label{sec:Intro}Introduction}
Ferroelectric materials show a variety of functional properties, which have potential for technological application. Examples are normal and inverse piezoelectric effects, which are used in pressure sensors, ultrasonic devices, or microactuators, the pyroelectric effect, which is used for thermal detectors, or the electrocaloric effect, which received significant attention for potential application in solid-state cooling \cite{Drag1998a, Setter2006, Martin2016}. These effects describe the dependence of the different state variables on the dielectric displacement $D$, which is particularly pronounced in ferroelectric materials due to the existence of a switchable permanent dipole moment resulting in a remanent polarization $P$. The specific dependence of the dielectric displacement on these state variables (such as, for example, the electric field $E$, stress $\sigma$, temperature $T$) is generally described by tensors of different rank. However, if one of these functional effects is targeted for applications, one also has to consider the interaction with the other state variables, which is often represented by so-called Heckmann diagrams \cite{Drag1998a, Lang1974b}. 

One example is the pyroelectric and the converse electrocaloric effect. It denotes a coupling of the dielectric displacement with temperature $T$ through the pyroelectric coefficient $p$. When all possible state variables are undetermined, the total value of the pyroelectric coefficient is composed mainly of four parts \cite{Jach2017a}: a primary pyroelectric effect caused by changes in spontaneous polarization, a secondary effect due to mechanical strain and the subsequent piezoelectric effect, a tertiary effect originating from spatial strain gradients, and a quaternary field-induced pyroelectric effect caused by the temperature dependence of the dielectric permittivity in the presence of an external electric field. We refer the reader to several excellent review articles \cite{Drag1998a, Jach2017a, Lubo2012a} and books \cite{Lang1974b} that elaborate on the thermodynamic background and the mathematical description of pyroelectricity. In general, the complex interplay of the state variables makes it difficult to disentangle the various sources of the pyroelectric response during standard measurements.

The situation becomes even more complicated if these effects are studied in thin films. In this case, the influence of the bulk substrate has to be taken into account, which results in mechanical boundary conditions because of the clamping at the interface. Additionally, the heat capacity of the substrate dominates the thermal properties of the sample and makes it difficult to measure temperature changes due to the electrocaloric effect. Therefore, alternative measurement strategies are of great interest in disengaging the influence of the different factors on the functional properties of ferroelectric materials.

In this paper, we exemplify a contactless measurement method which might enable us to distinguish between different contributions to the pyroelectric coefficient in thin ferroelectric films. We chose (Ba,Sr)TiO$_{3}$ as model system, which was used for electrocaloric studies before \cite{Magalhaes2021}. Our measurement relies on impulsive heating of the interface of an ferroelectric thin film and a metallic bottom electrode layer with ultra-short light pulses. Subsequently, we probe the sample temperature by time-resolved x-ray diffraction (TR-XRD) \cite{shay2020,navi2014a,shay2011}. In contrast to earlier approaches \cite{Chyn1956a}, we measure the transient sample temperature in the thin film system, thereby determining an additional deformation that relates to the pyroelectric response.

\section{\label{sec:PyroResponse} Analysis of the pyroelectric coefficient by hysteresis loop measurements}
The pyroelectric coefficient in ferroelectric thin films can be determined by measuring field-induced non-linear polarization hysteresis loops at different temperatures. We specifically investigated the properties of epitaxial Ba$_{0.7}$Sr$_{0.3}$TiO$_{3}$ films grown on NdScO$_{3}$. Details on the preparation and basic characterization of these samples were published previously \cite{Wawra2023}. 

\begin{figure}[htbp]
	\centering
	\includegraphics[width = \columnwidth]{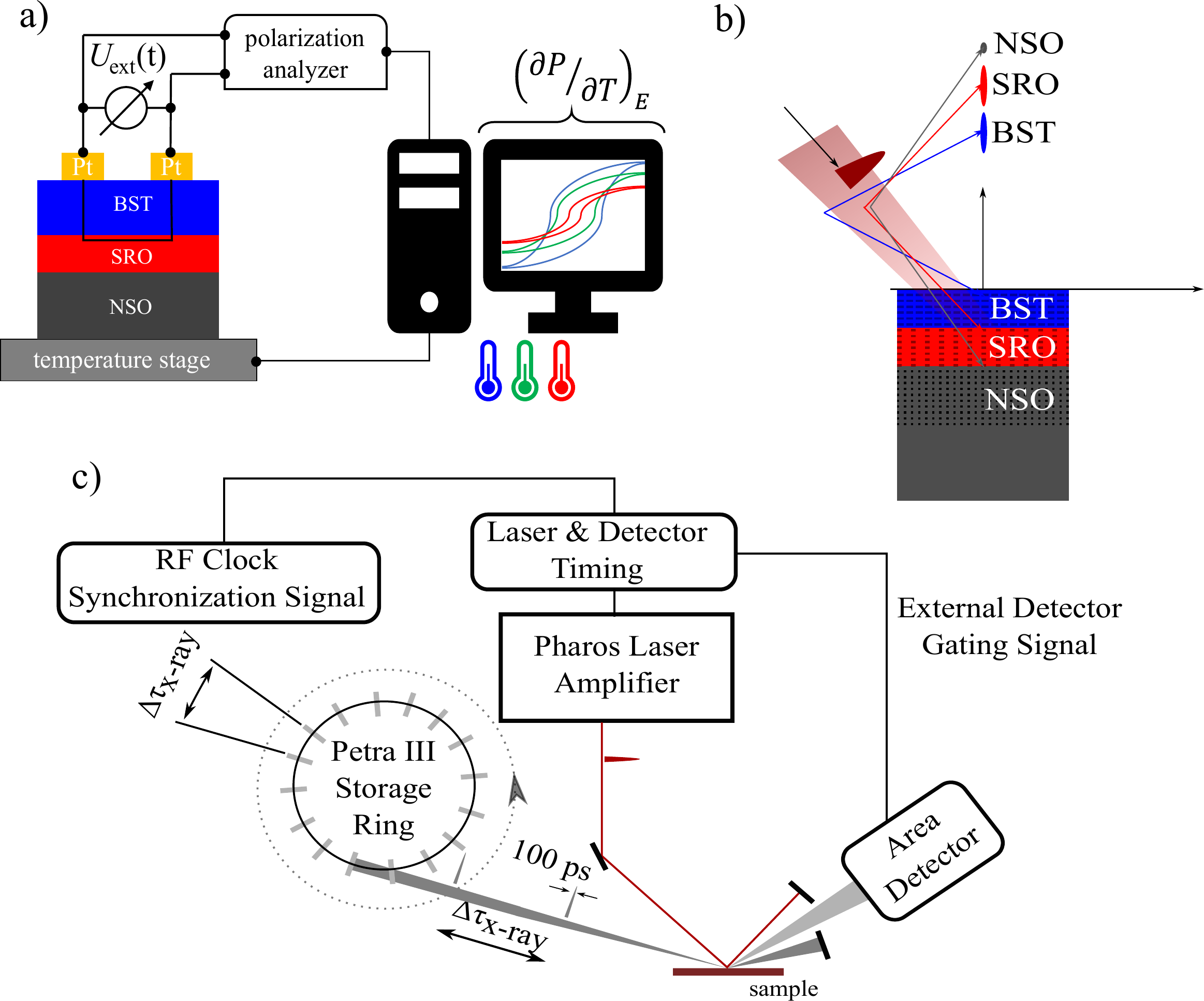}
	\caption{\textbf{Experimental Method:} a) Experimental setup for measuring polarization hysteresis loops to derive the pyroelectric and EC effect. b) Optical excitation and subsequent XRD probing of the out-of-plane lattice expansion in symmetric Bragg geometry. BST, SRO and NGO denote the film materials (Ba,Sr)TiO$_{3}$, SrRuO$_{3}$ and the NdGaO$_{3}$ substrate. c) Experimental setup for optical x-ray pump-probe measurements. For details see the main text and the supplementary information.}
	\label{fig:Setups}
\end{figure}

Figure~\ref{fig:Setups}~a) depicts our experimental setup where we perform $P(E,T)$ polarization measurements. The sample is a 250\,nm thin (Ba,Sr)TiO$_{3}$ epitaxial film grown on a NdScO$_{3}$ substrate with an intermediate 20\,nm SrRuO$_{3}$ bottom electrode. It is mounted on a temperature stage to set temperatures from -130\,$^{\circ}$C to 130\,$^{\circ}$C. The top electrodes are connected to a polarization analyzer that applies the electric field and measures the dielectric polarization in the thin film in out-of-plane direction. The measurement is controlled by software to loop the external electric field at different sample temperatures with a frequency of $f=$1\,kHz. Slight inaccuracies of the measurement device make it necessary to fit the $P(E)$ curves to precisely determine $P(T)$ at exact values of the electric field $E$ and consequently $p = (\partial P / \partial T)_{E}$. We analyze the upper positive branch of the hysteresis loops only to minimize ferroelectric domain switching contributions to the polarization change. Still, the acquired values for $p$ consist of the sum of the primary pyroelectric coefficient and the higher-order terms mentioned above \cite{Jach2017a}.

\begin{figure}[htbp]
	\centering
	\includegraphics[width = \columnwidth]{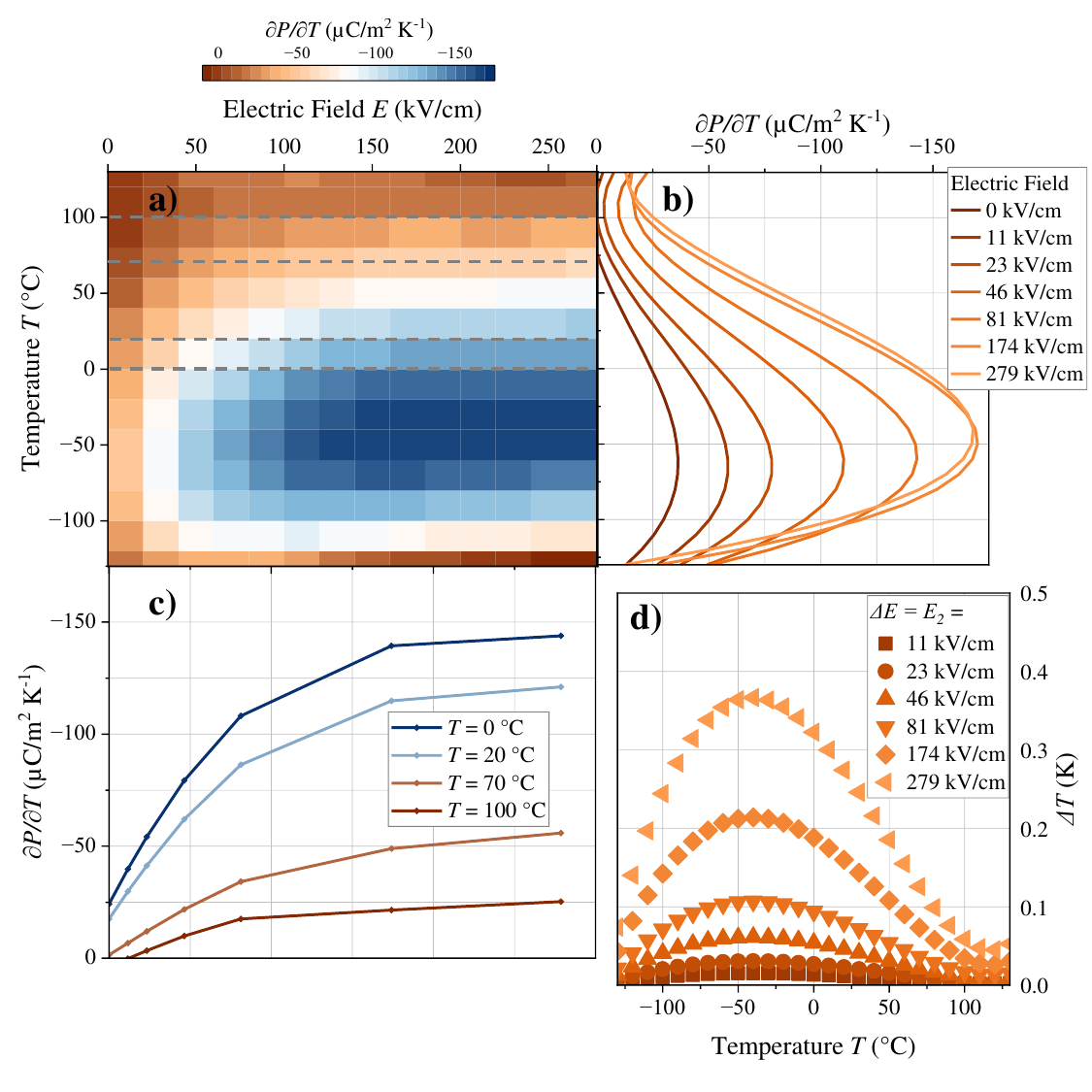}
	\caption{\textbf{Characterization of the pyroelectric effect from polarization measurements:} a) Full dataset depicting the dependence of the pyroelectric coefficient on external electric field and temperature. b) Dependence of pyroelectric coefficient on temperature for selected electric fields extracted from the data shown in a). The coefficient increases with increasing electric field.  c) Field-dependent pyroelectric effect at 0\,$^{\circ}$C, 20\,$^{\circ}$C, 70\,$^{\circ}$C and 100\,$^{\circ}$C. d) Electrocaloric temperature change derived from data using the indirect method.}
	\label{fig:Indirect}
\end{figure}

The complete dataset obtained in our measurement is depicted in Figure~\ref{fig:Indirect}~a). More specifically, the values for the pyroelectric coefficient $p$ are obtained from the derivative of the $P(T)$ fits at different electric fields $E$. As an example, seven of those $p(T) = (\partial P / \partial T)_{E} (T)$ curves are shown in Fig~\ref{fig:Indirect}~b) between 0 and the maximum applied field of of 279\,kV/cm. Reading the values at constant temperatures allows us to obtain the dependence of $p(E) = (\partial P / \partial T)_{E} (E)$ on the electric field. Figure~\ref{fig:Indirect}~c) shows this dependence of the pyroelectric coefficient at four different temperatures, namely 0\,$^{\circ}$C (dark blue), 20\,$^{\circ}$C (light blue), 70\,$^{\circ}$C (orange) and 100\,$^{\circ}$C (brown). Dashed lines in Figure~\ref{fig:Indirect}~a) denote the cuts in the full dataset. The pyroelectric coefficient increases with the external electric field and reaches a maximum of about -170\,$\mu$C\,m$^{-2}$\,K$^{-1}$ at temperatures around -50\,$^{\circ}$C. This value is only slightly lower than the pyroelectric coefficient of -200\,$\mu$C\,m$^{-2}$\,K$^{-1}$ measured on bulk BaTiO$_{3}$ \cite{Bowen2014} as well as on Ba$_{0.6}$Sr$_{0.4}$TiO$_{3}$ films grown on SrTiO$_{3}$ substrates \cite{Tong2014}. It slowly decreases towards higher temperatures. Usually, the highest pyroelectric coefficient is expected close to the phase transition temperature and is accompanied by a strong dielectric response. For bulk Ba$_{0.7}$Sr$_{0.3}$TiO$_{3}$ this transition occurs close to room temperature \cite{Acosta2017, Vendik2000}. In our sample, $p$ gradually changes in this temperature range, indicating a diffuse transition as is typical for clamped thin films and solid solution materials. Furthermore, tensile strain induced by the substrate shifts the maximum dielectric response of the ferroelectric film to lower temperatures \cite{Schl2014a, Schm2017a, schw2017a}. Theoretical calculations for Ba$_{0.7}$Sr$_{0.3}$TiO$_{3}$ predict for tensile strain a sequence from a paraelectric over a completely in-plane polarized $aa$ to a complex in-plane and out-of-plane polarized $r$ phase, where the first transition is above and the second below room temperature, respectively \cite{Ban2002}. As we probe only the out-of-plane component of the polarization in our measurement configuration, we find a strong polarization change at the assumed transition between the $aa$ and the $r$ phase below room temperature, which explains the highest pyroelectric coefficient at about -50°C. 

For further characterization, we also determined the electrocaloric response of our film. As mentioned above, the electrocaloric effect is the inverse of the pyroelectric effect. It describes the adiabatic temperature change due to an alternating electric field $\Delta E$. The field causes the dipoles to rearrange, alternating the entropy of the polar subsystem. Changes in the thermal subsystem compensate for this with the electrocaloric temperature change $\Delta T$.

In bulk materials, calorimetry methods are used to measure the temperature change directly \cite{Liu2016}. Unfortunately, this becomes unfeasible for thin ($<\mu$m) or ultra thin ($<$100 \,nm) films due to the small active volume and the overshadowing influence of the substrate. Accordingly, the quantification of the electrocaloric effect is typically done by the so-called indirect method, which is often the only possibility. It deduces the entropy change due to the change in electric field via the thermodynamic Maxwell relation from the pyroelectric coefficient $p$:
\begin{align}
    (\partial S/\partial E)_{T} = (\partial P / \partial T)_{{E}} = p
    \label{equ:PyroCoeffIndirect}
\end{align}

 The temperature change can be calculated by \cite{kutnjak2015i}:
 
\begin{align}
        \Delta T = -\int_{{E}_{1}}^{{E}_{2}} \frac{T}{\rho C_{p}({E},T)}\left( \frac{\partial P}{\partial T}\right)_{E} dE
    \label{equ:ECbasic}
\end{align}
where $E_2-E_1=\Delta E$ relate to the change of the electric field and $\rho$ and $C_{p}$ denote the mass density and specific heat, respectively, which can be derived from literature values \cite{strukov2003}.

Consequently, our measured values for $p$ can be utilized to determine $\Delta T$ for the withdrawal of different electric fields. The results are shown in Figure~\ref{fig:Indirect}~d). For $E_2$ the same fields as shown in b) are used and $E_1$ was set to zero. The maximum $\Delta T$ was determined at temperatures around -50\,$^{\circ}$C and for the withdrawal of the maximum electric field. In this case, a temperature change of 0.37\,K is calculated. Note that all different contributions from the pyroelectric effect contribute to the electrocaloric effect as well and that the indirect method does not discriminate between different orders of $p$.

\section{Time-resolved x-ray diffraction}\label{sec:trXRD}
The secondary pyroelectric effect is related to mechanical strain resulting in an additional change of polarization due to the piezoelectric effect. To quantify this contribution, the dependence of the mechanical strain on temperature needs to be known for our thin layers. Therefore, we performed time-resolved x-ray diffraction measurements to determine the deformation dynamics of (Ba,Sr)TiO$_{3}$ after excitation with femtosecond laser pulses. For this experiment we use a 100\,nm thin Ba$_{0.7}$Sr$_{0.3}$TiO$_{3}$ film which is grown on a SrRuO$_{3}$ electrode on top of a (110) oriented orthorhombic NdScO$_{3}$ substrate. The orientation of both thin films is the (001) direction in the pseudocubic notation of the substrate. With the exception of the (Ba,Sr)TiO$_{3}$ film thickness, the sample stack is identical to the sample used in the polarization loop measurement. We prefer a thinner (Ba,Sr)TiO$_{3}$ layer to reach a homogeneous temperature profile in earlier pump-probe delays.

The time-resolved experiment is depicted in Figure~\ref{fig:Setups}~b) and~c). We employ a pump-probe scheme, where an intense ultrashort light pulse is absorbed into the SrRuO$_{3}$ layer of the sample. The (Ba,Sr)TiO$_{3}$ layer and the substrate are transparent at the wavelength of the excitation laser of 1030\,nm. At every delay point we separately scanned the incidence angle of the x-ray beam around the SrRuO$_{3}$ and (Ba,Sr)TiO$_{3}$ (002)$_{pc}$ Bragg reflection. We converted the diffraction data to reciprocal space using the xrayutilities python package \cite{Kriegner2025} to obtain time-dependent reciprocal space maps (RSMs) \cite{schi2013a}, which were further analyzed to obtain transient positions of the Bragg reflex and thus the transient lattice parameter. The sample was kept at room temperature at negative time delay and we derive the average transient temperature of each layer from the observed  shift of the (002)$_{pc}$ Bragg reflection via the relation \cite{shay2011}
\begin{equation}
    \Delta q_{z} = \frac{q_{z,0}}{\alpha\Delta T +1}.
    \label{equ:QToTemp}
\end{equation}
We performed the measurements at the P08 beamline of the PETRA III storage ring at DESY. The photon energy was tuned to 9\,keV by a double crystal monochromator in the beam path. The relative energy bandwidth at P08 in monochromatic mode is 10$^{-4}$ and the reciprocal space resolution of our scans is 2\,$\mu$m$^{-1}$. X-ray pulses with a duration of 120\,ps impinge the sample at a repetition rate of 5.2 MHz, i.e., every 192 \,ns. We adapted the rate of x-ray pulses detected in the measurement to the arrival rate of the laser pump pulses by applying an external gating signal to the x-ray detector. The beamline was equipped with a femtosecond laser amplifier (Pharos, Light Conversion) that emitted optical pulses with a duration of 290\,fs at a wavelength of 1030\,nm. The laser was synchronized with the RF clock of the storage ring to control the pump-probe delay. We performed the measurement at a repetition rate of 10\,kHz. The area cross section of the laser beam was $\pi\times(250\,\mu m\times 300\,\mu m)$ and we accounted for the incidence angle on the sample to calculate the laser fluence. The laser was p-polarized with respect to the diffraction plane to minimize reflection losses.

\begin{figure}[htbp]
	\centering
	\includegraphics[width = \columnwidth]{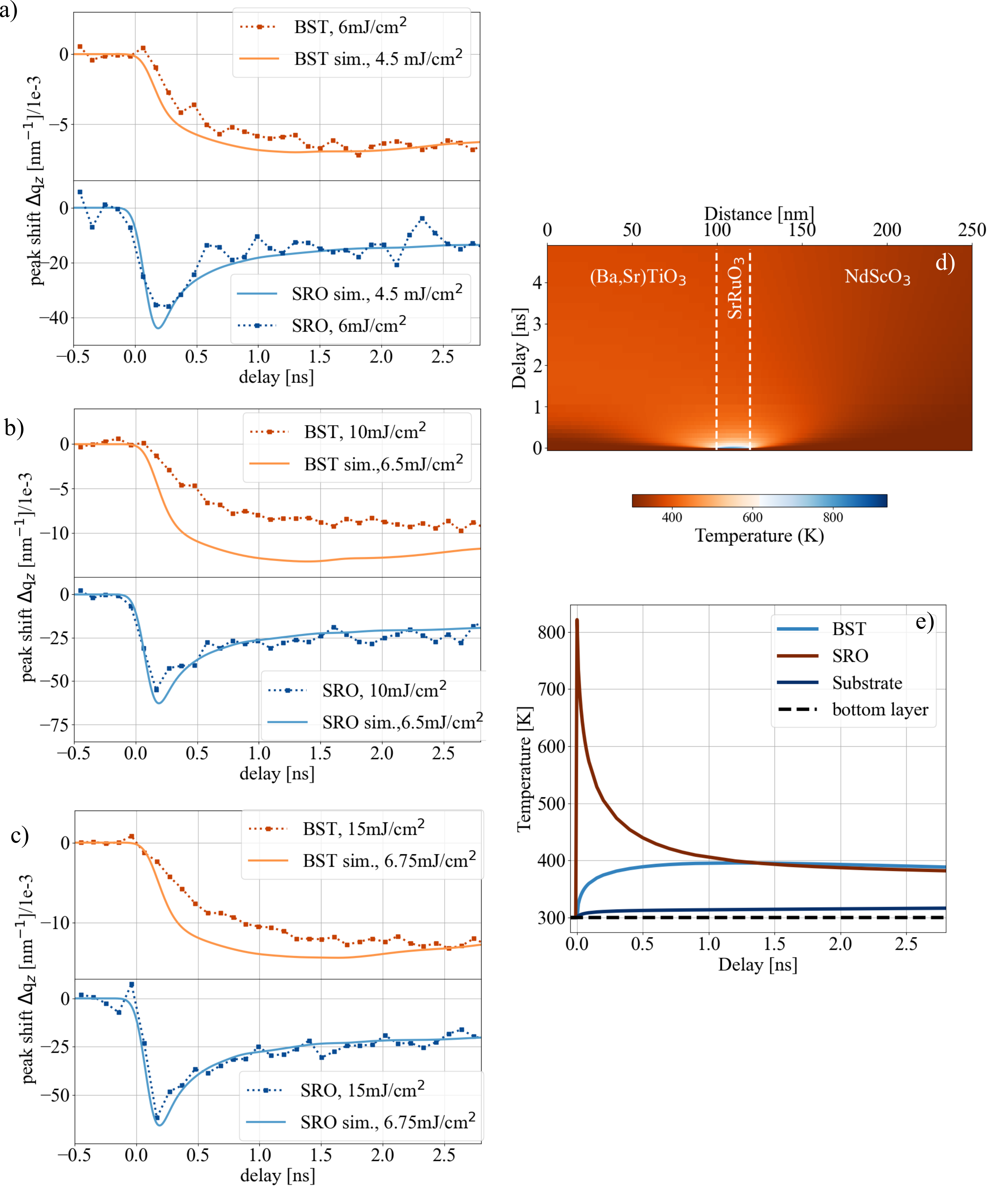}
	\caption{\textbf{Results of time-resolved measurements:} a)-c) Measured (symbols) and simulated (solid lines) shift of the symmetric (002) Bragg reflex of the (Ba,Sr)TiO$_{3}$ (upper, red) and SrRuO$_{3}$ (lower, blue) layer. The simulation result is convoluted with a gaussian function of full width at half maximum of 120\,ps to account for the limited time resolution of the experiment. d) Transient sample temperature plotted along the surface normal. The interface between the layers and the substrate are indicated by white dashed lines. The simulation was performed for a fluence of 6.5\,mJ/cm$^{2}$ e) Transient average temperature in ferroelectric (Ba,Sr)TiO$_{3}$, SrRuO$_{3}$ and in the substrate calculated from the data depicted in d).}
    \label{fig:ExpDirect}
\end{figure}

Figure~\ref{fig:ExpDirect}~a) to c) depict the transient shift of the (002) Bragg peak of (Ba,Sr)TiO$_{3}$ (upper, red) and SrRuO$_{3}$ (lower, blue) layer in reciprocal space coordinate $q_{z}$ after optical excitation with a fluence of 6, 10 and 15\,mJ/cm$^{2}$, respectively. The square symbols represent the measurement results, and the solid lines (Fig.~\ref{fig:ExpDirect}~ b) and c)) denote heat diffusion simulations. The simulated curves were convoluted with a Gaussian function (full width at half maximum = 120\,ps) to account for the limited time resolution of the experiment.

To simulate heat diffusion dynamics, we employ the udkm1Dsim Python toolbox \cite{Schi2021a}. The package solves the heat diffusion equation in the thin film heterostructure after optical excitation \cite{shay2020} using a finite element method. Specifically, we solve the heat diffusion equation for a time-dependent, delta-like excitation, i.e., the 290 fs short laser pulse which is only absorbed in the 20 nm thin SrRuO$_3$ electrode. The laser footprint has a lateral radius of more than 100\,$\mu$m. These dimensions determine the direction of the heat flow: The thermal gradient along the surface normal is $\Delta$T$_{\text{laser}}$/d$_{SRO}$ = 5\,K/nm which is three orders of magnitude stronger than the thermal gradient in in-plane direction $\Delta$T$_{\text{laser}}$/d$_{\text{footprint}}$ = 1$\cdot$10$^{-3}$\,K/nm. Because heat diffuses in the direction of the thermal gradient, we neglect the in-plane heat diffusion for the first 10 ns after the laser excitation \cite{navi2014a}. Input parameters for the simulation are the thermal expansion coefficient $\alpha$, the specific heat c$_p$, the thermal conductivity $k$ and the refractive index $n$. Most values can be found in literature and are listed in the supplement. We measured the thermal expansion coefficient for (Ba,Sr)TiO$_3$ of our samples using temperature-dependent XRD (see supplement). We used a refractive index of SrRuO$_3$ of 2.44+4.32j \cite{Schi2021a,Kost1998a}. It is important to point out, that the simulated transient temperature in a specific layer, e.g., SrRuO$_3$ also depends on the material parameters of the adjacent layers. Due to the agreement of the measured and simulated transients of the SrRuO$_3$ layer, we have high confidence in our simulation model. In addition, we applied this model already to interpret data from earlier experiments \cite{shay2020,shay2011,Schi2021a,Gaal2023a}.

As an example, we depict the simulated temperature along the surface normal in figure~\ref{fig:ExpDirect}~d) and the transient average temperature in each layer and in the substrate in figure~\ref{fig:ExpDirect}~e). The simulation was performed for a fluence of 6.5\,mJ/cm$^{2}$.
To complete our modeling of the experiment, we calculate the x-ray diffraction efficiency from dynamical diffraction theory \cite{warr1990a} and determine the center-of-mass coordinate $q_{z}$ of the (Ba,Sr)TiO$_{3}$ and SrRuO$_{3}$ Bragg peak. We use the same numerical method to extract the peak position from the experimental and simulation data. Details of the simulation model are given in the supplemental material.

We again stress that the laser pulses were only absorbed by free electrons in the metallic SrRuO$_{3}$ electrode. The excited hot electrons thermalize by electron-phonon coupling with a well-known electron-phonon coupling time constant in SrRuO$_{3}$ of less than 500\,fs \cite{Matt2021a, Bohj2012a}. This leads to an impulsive thermal expansion of the SrRuO$_{3}$ film. Subsequently, the heat flows into the adjacent (Ba,Sr)TiO$_{3}$ layer and the substrate. We employ the SrRuO$_{3}$ film as a temperature sensor by converting the deformation to a temperature via the thermal expansion coefficient. In contrast, it is not easy to convert the measured deformation of (Ba,Sr)TiO$_{3}$ to a temperature due to its piezoelectric response. In the following, we discuss the heat diffusion dynamics in the sample for three different excitation fluences.

First, we analyze the measurement with a low excitation fluence of 6\,mJ/cm$^{2}$ (c.f. Figure~\ref{fig:ExpDirect}~a)). With the limited time resolution of the experiment of 120\,ps we observe an immediate shift of the (002) SrRuO$_{3}$ Bragg peak depicted in Figure~\ref{fig:ExpDirect}~a) (lower, blue symbols). Subsequently, the temperature in the adjacent (Ba,Sr)TiO$_{3}$ layer increases by heat diffusion, again shifting the (002) (Ba,Sr)TiO$_{3}$ Bragg peak (upper, orange symbols) to lower angles due to thermal expansion of the lattice. We compare the measured data to thermal expansion that results from heat diffusion simulations which yield the temperatures in the top (Ba,Sr)TiO$_{3}$ film and in the SrRuO$_{3}$ electrode. With an excitation fluence of 4.5\,mJ/cm$^{2}$ (solid lines), the simulation quantitatively reproduces the measured deformation (dotted lines and bullets) in both films.  

Second, we discuss a measurement in which we excited the sample with a fluence of 10\,mJ/cm$^{2}$. The measured data (bullets) and the simulation results (solid lines) are depicted in Figure~\ref{fig:ExpDirect}~b). Again, we observe an impulsive and a gradual temperature rise in SrRuO$_{3}$ and (Ba,Sr)TiO$_{3}$, respectively. Our model reproduces the expansion dynamics in the SrRuO$_{3}$ temperature sensor for an excitation fluence of 65\% of the measured value. Unlike the first measurement, the simple heat diffusion model clearly overestimates the expansion of (Ba,Sr)TiO$_{3}$.

Third, we discuss a measurement with an excitation fluence of 15\,mJ/cm$^{2}$ in Figure~\ref{fig:ExpDirect}~c). Again, we find an overlap of the simulated and measured expansion in the SrRuO$_{3}$ temperature sensor layer; however, the simulated fluence amounts only to 45\% of the fluence value set in the first experiment. The deformation of (Ba,Sr)TiO$_{3}$ does not comply with the predictions of the heat diffusion model and again shows a deviation as in the second measurement.

In summary, our measurements show that we can effectively employ a SrRuO$_{3}$ thin film as temperature sensor. The temperature within the heterostructure is equilibrated 1.5\,ns after the excitation pulse is absorbed. At higher pump fluence the energy transfer from the exciting laser pulse to the lattice temperature is reduced. We account for this effect by reducing the excitation fluence to 65\% and 45\% for the 10\,mJ/cm$^{2}$ and 15\,mJ/cm$^{2}$ measurements, respectively. At the same time, the deformation of the (Ba,Sr)TiO$_{3}$ film does not follow the prediction of the heat diffusion model. We propose a charge screening mechanism at the (Ba,Sr)TiO$_{3}$ to SrRuO$_{3}$ interface, which can explain our observations. More details will be discussed in the next section.

\section{Dynamics of the Ferroelectric Polarization after excitation with femtosecond laser pulses}
\label{sec:FEdynamics}
We now attempt an interpretation of the experimental data presented in the previous section. In particular, we focus on the deviation of the measured and simulated strain in our sample. Because of the good agreement in the low-fluence measurement we consider this to be a calibration of the material parameters and the simulation model. At higher fluences the measured deformation in both films is less pronounced than predicted by the simulation. This suggests the existence of an energy dissipation channel that influences the observed deformation dynamics above a certain excitation threshold. We did not observe a threshold behavior in the polarization loop measurements in section~\ref{sec:PyroResponse}. Therefore, we believe that the dissipation occurs only in a non-equilibrium state of the sample. 

\begin{figure}[htbp]
	\centering
	\includegraphics[width = \columnwidth]{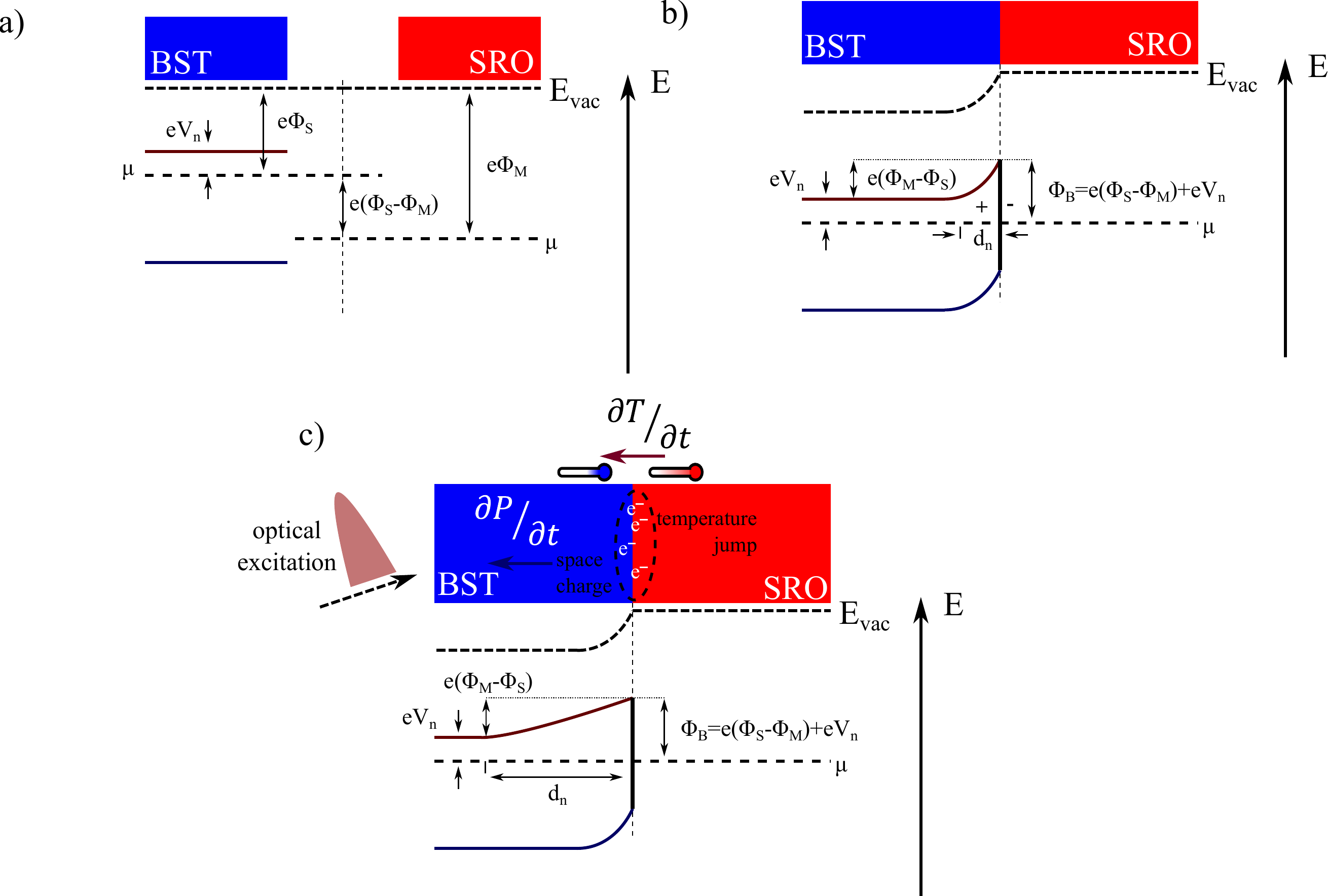}
	\caption{\textbf{Electronic properties of the (Ba,Sr)TiO$_{3}$/SrRuO$_{3}$ interface:} a) Band structure of (Ba,Sr)TiO$_{3}$ and SrRuO$_{3}$ showing an offset in the Fermi energy of $\mu$ in both materials. b) Band structure of (Ba,Sr)TiO$_{3}$ grown on an SrRuO$_{3}$ electrode. The sketch depicts the formation of a Schottky barrier with a height of $\phi_{B}=e(\phi_{M}-\phi_{S})+eV_{n}$. c) Dynamics at the interface after absorption of a femtosecond laser pulse in SrRuO$_{3}$. The optical excitation leads to an injection of charges from SrRuO$_{3}$ into (Ba,Sr)TiO$_{3}$ across the Schottky barrier and a temperature jump in SrRuO$_{3}$ followed by thermal diffusion into (Ba,Sr)TiO$_{3}$.}
	\label{fig:Interp}
\end{figure}

We start our discussion by introducing the band structure of the individual films and of the interface. It is depicted in Figure~\ref{fig:Interp}~a) and b), respectively. Due to the difference of the work functions $\phi_{S}$ in (Ba,Sr)TiO$_{3}$ and $\phi_{M}$ in SrRuO$_{3}$, the interface forms a Schottky barrier with a height of $\phi_{B}=e(\phi_{M}-\phi_{S})+eV_{n}$, where $V_{n}$ is the energy offset of the conduction band from the Fermi energy $\mu$.  The characteristic of the Schottky barrier is the space charge region, which extends into the (Ba,Sr)TiO$_{3}$ film. It results from the accumulation of charges at the (Ba,Sr)TiO$_{3}$/SrRuO$_{3}$ interface. Its extension $d_n$ into (Ba,Sr)TiO$_{3}$ increases with decreasing space charge density n$_{D}$ \cite{Hubm2016a, gros2014a}:
\begin{equation}
    d_n=\left(\frac{2\epsilon_{r}\epsilon_{0}}{e n_{D}}(\phi_{M}-\phi_{S}) \right)^{1/2}
    \label{equ:spacecharge}
\end{equation}
Literature values for the work function in SrRu0$_{3}$ $\phi_{m}$ range from 4.6\,eV to 4.9\,eV \cite{Hart2000a}. Several sources discuss the permeability in (Ba,Sr)TiO$_{3}$ and relate the electronic and structural properties \cite{Free2016a,Lee2016a} or report on the influences of the Schottky barrier \cite{Xi2017a}. However, to our knowledge, no quantitative values are reported.

We now consider how the band structure changes in a transient state after the excitation pulse is absorbed in the SrRuO$_{3}$ film. The situation is depicted in Figure~\ref{fig:Interp}~c). Initially, the laser is absorbed by the electrons in the conduction band in SrRuO$_{3}$~\cite{wei2017a,Guo2021a}. Due to the low specific heat, which is characterized by a Sommerfeld coefficient of $\approx$0.13\,J/(K$^2$kg), the electron temperature rises to extreme values of more than 2000\,K. At the resulting thermal kinetic energies the Schottky barrier becomes negligible and electrons can penetrate across the interface and into the (Ba,Sr)TiO$_{3}$ film. Similar ballistic energy transport across an interface in nanoscale systems has also been observed elsewhere \cite{Plec2024a, Herz2022a}. The free electrons in (Ba,Sr)TiO$_{3}$ shield the positive space charge density $n_{D}$, thus extending the space charge region further away from the interface and towards the sample surface. As a result the band structure deforms in a significant volume of the (Ba,Sr)TiO$_{3}$ film, which effectively changes the electric field $E$ in the space charge region of (Ba,Sr)TiO$_{3}$.

The impact of the optical excitation at the Schottky barrier on the pyroelectric properties might be twofold. First, the transient change of the space-charge region is accompanied by the formation of an electric field. This results in additional strain due to the inverse piezoelectric effect and therefore counteracts the thermal expansion due to the laser-induced temperature change. We observe this as a frustrated deformation in our time-resolved XRD measurement. Second, the injection of free electrons into the conduction band of (Ba,Sr)TiO$_{3}$ results in losses and thereby changes the permittivity of the materials as well as the resulting polarization. 

To prove our assumption, one needs a better characterization of the electronic properties of the Schottky contact. One possibility to obtain this information is Deep Level Transient Spectroscopy (DLTS) \cite{Lang1974a,Stol1985a}. This method measures the evolution of the interface capacity after injecting charges across the Schottky barrier. Thus, it mimics the mechanism that is presumably responsible for our observed changes of the electric field in the space-charge region. 

\section{Comparison of time-resolved deformation and polarization loop measurements}

In this final section, we want to discuss how the results of the time-resolved deformation measurement in section~\ref{sec:trXRD} can give additional information on the pyroelectric coefficient, which was determined from the hysteresis loop measurement in section~\ref{sec:PyroResponse}. In general, the pyroelectric effect describes the change of polarization with temperature, which is typically measured as a change of surface charges in the film architecture. As already mentioned in the introduction, different mechanisms contribute to the pyroelectric response. The polarization measurements in section~\ref{sec:PyroResponse} cannot distinguish between the different contributions; therefore, only the sum of all parts is measured.  In contrast, the measurements in section~\ref{sec:trXRD} probe the structural changes with temperatures. The resulting strain contributes mainly to the secondary (related to the piezoelectric effect) and tertiary (related to strain gradients) pyroelectric effect. As we assume a homogeneous strain state at 1.5\,ns after the laser pulse due to a homogeneous temperature distribution in the (Ba,Sr)TiO$_{3}$ layer, the tertiary pyroelectric contribution can be neglected. Therefore, the local deformation allows us to calculate the secondary pyroelectric effect. In general, the following relationship is given for this contribution \cite{Jach2017a}:
\begin{equation}
    {p_{sec}} = \sum_{i,j,k}{d_{ij}c_{jk}\alpha_k} 
    \label{equ:pyro2nd}
\end{equation}
where $d_{ij}$ denotes the piezoelectric stress tensor at constant temperature, $c_{jk}$ the elastic stiffness tensor at constant temperature and $\alpha_{k}$ the thermal expansion coefficient, respectively. 
On the time scale of our experiment, heat diffusion within the sample and the subsequent thermal expansion occur only along the out-of-plane direction. Whereas the lattice change is measured by XRD, the corresponding temperature change can be taken from the thermal simulations. To determine the value of the secondary piezoelectric effect one needs to take into account additionally the piezoelectric and elastic stiffness constants. Calculations for BaTiO$_{3}$ using the Devonshire theory show a clear dependence of these parameters on temperature as well as on crystal structure \cite{Acosta2017}. As the temperature change is in the order of 100 K for our experiment (compare Fig.~\ref{fig:ExpDirect}~e)), the temperature dependence cannot be neglected. Furthermore, detailed structural investigations of our Ba$_{0.7}$Sr$_{0.3}$TiO$_{3}$ grown on NdScO$_{3}$ indicate a distorted tetragonal or orthorhombic structure due to a fully strained growth with a shorter lattice parameter in the out-of-plane direction \cite{Wawra2023}. Considering the unknown temperature dependence of the piezoelectric and elastic stiffness of our compound as well as the uncertainty of the crystal symmetry, no quantitative assessment of the secondary pyroelectric effect can be made currently. In addition, possible changes of the electronic properties at the boundary between SrRuO$_{3}$ and (Ba,Sr)TiO$_{3}$ as discussed in section~\ref{sec:FEdynamics} will result in additional losses and simultaneously in a change of the dielectric permittivity, which will also influence the pyroelectric effect. To quantitatively assess the individual components of the pyroelectric coefficient, we will attempt a better, i.e. temperature-dependent characterization of the relevant material parameters of our samples in the future. 

\section{Conclusion}
\label{sec:conclusion}
In conclusion, we have determined the pyroelectric coefficient of an epitaxial (Ba,Sr)TiO$_{3}$ thin from temperature-dependent polarization measurements. Additionally, we used time-resolved x-ray diffraction measurements to obtain more insight in the deformation dynamics of our ferroelectric films. The time-resolved experiment employs femtosecond laser pulses to excite a (Ba,Sr)TiO$_{3}$/SrRuO$_{3}$ heterostructure. Comparison with heat diffusion simulations reveals a frustration of the thermal expansion. The reduced thermal expansion indicates a charge dissipation channel from the laser-excited SrRuO$_{3}$ across the Schottky barrier into (Ba,Sr)TiO$_{3}$, which can occur only in a transient state. So far we were unable to achieve a quantitative comparison of the static and transient experiments due to an incomplete characterization of material parameters such as the temperature dependence of the piezoelectric coefficients or the charge dynamics at the Schottky barrier after laser excitation. A modified sample design will facilitate the characterization of these parameters in the future.  

\section*{Acknowledgment}
We gratefully acknowledge funding by the Deutsche Forschungsgemeinschaft under DFG grants GA2558/5 and HU1726/8. The authors thank Michael Kühnel for technical support, as well as Christian Molin and Sylvia E. Gebhardt (Fraunhofer IKTS Dresden) for the provision of the (Ba,Sr)TiO$_{3}$ target. We acknowledge DESY (Hamburg, Germany), a member of the Helmholtz Association HGF, for the provision of experimental facilities. Parts of this research were carried out at PETRA III and we would like to thank Florian Betram for assistance in using beamline P08. Beamtime was allocated for proposal I-20231316. We also thank Andreas Fiedler and Martin Schmidbauer (both Institute für Kristallz{\"u}chtung IKZ) for fruitful discussions.

\section*{Data Availability Statement}
All data in this work are available on request through contact with the
corresponding author.

\appendix



%

\bibliography{bibcollection}

\begin{thebibliography}{10}
\expandafter\ifx\csname url\endcsname\relax
  \def\url#1{\texttt{#1}}\fi
\expandafter\ifx\csname urlprefix\endcsname\relax\def\urlprefix{URL }\fi
\expandafter\ifx\csname href\endcsname\relax
  \def\href#1#2{#2} \def\path#1{#1}\fi

\bibitem{Drag1998a}
D.~Damjanovic, Ferroelectric, dielectric and piezoelectric properties of
  ferroelectric thin films and ceramics, Reports on Progress in Physics 61~(9)
  (1998) 1267.
\newblock \href {http://dx.doi.org/10.1088/0034-4885/61/9/002}
  {\path{doi:10.1088/0034-4885/61/9/002}}.

\bibitem{Setter2006}
N.~Setter, D.~Damjanovic, L.~Eng, G.~Fox, S.~Gevorgian, S.~Hong, A.~Kingon,
  H.~Kohlstedt, N.~Y. Park, G.~B. Stephenson, I.~Stolitchnov, A.~K. Taganstev,
  D.~V. Taylor, T.~Yamada, S.~Streiffer, Ferroelectric thin films: review of
  materials, properties, and applications, Journal of Applied Physics 100~(5)
  (2006) 051606.
\newblock \href {http://dx.doi.org/10.1063/1.2336999}
  {\path{doi:10.1063/1.2336999}}.

\bibitem{Martin2016}
L.~W. Martin, A.~M. Rappe, Thin-film ferroelectric materials and their
  applications, Nature Review Materials 2~(2) (2016) 16087.
\newblock \href {http://dx.doi.org/10.1038/natrevmats.2016.87}
  {\path{doi:10.1038/natrevmats.2016.87}}.

\bibitem{Lang1974b}
S.~B. Lang, Sourcebook of Pyroelectricity, CRC Press, Boca Raton, FL, 1974.

\bibitem{Jach2017a}
S.~Jachalke, E.~Mehner, H.~Stöcker, J.~Hanzig, M.~Sonntag, T.~Weigel,
  T.~Leisegang, D.~C. Meyer, How to measure the pyroelectric coefficient?,
  Applied Physics Reviews 4~(2) (2017) 021303.
\newblock \href {http://dx.doi.org/10.1063/1.4983118}
  {\path{doi:10.1063/1.4983118}}.

\bibitem{Lubo2012a}
I.~Lubomirsky, O.~Stafsudd, Invited review article: Practical guide for
  pyroelectric measurements, Review of Scientific Instruments 83~(5) (2012)
  051101.
\newblock \href {http://dx.doi.org/10.1063/1.4709621}
  {\path{doi:10.1063/1.4709621}}.

\bibitem{Magalhaes2021}
B.~Magalhaes, S.~Engelhardt, C.~Molin, S.~Gebhardt, K.~Nielsch, R.~Hühne,
  Electrocaloric temperature changes in epitaxial {Ba$_{1-x}$Sr$_x$TiO$_3$}
  films, Journal of Alloys and Compounds 891 (2021) 162041.
\newblock \href {http://dx.doi.org/10.1016/j.jallcom.2021.162041}
  {\path{doi:10.1016/j.jallcom.2021.162041}}.

\bibitem{shay2020}
R.~Shayduk, P.~Gaal, Transition regime in the ultrafast laser heating of
  solids, Journal of Applied Physics 127~(7) (2020) 073101.
\newblock \href {http://dx.doi.org/10.1063/1.5143717}
  {\path{doi:10.1063/1.5143717}}.

\bibitem{navi2014a}
H.~A. Navirian, D.~Schick, P.~Gaal, W.~Leitenberger, R.~Shayduk, M.~Bargheer,
  Thermoelastic study of nanolayered structures using time-resolved x-ray
  diffraction at high repetition rate, Applied Physics Letters 104~(2) (2014)
  021906.
\newblock \href {http://dx.doi.org/10.1063/1.4861873}
  {\path{doi:10.1063/1.4861873}}.

\bibitem{shay2011}
R.~Shayduk, H.~Navirian, W.~Leitenberger, J.~Goldshteyn, I.~Vrejoiu,
  M.~Weinelt, P.~Gaal, M.~Herzog, C.~V.~K. Schmising, M.~Bargheer, {Nanoscale
  heat transport studied by high-resolution time-resolved x-ray diffraction},
  New J. Phys. 13~(9) (2011) 093032.
\newblock \href {http://dx.doi.org/10.1088/1367-2630/13/9/093032}
  {\path{doi:10.1088/1367-2630/13/9/093032}}.

\bibitem{Chyn1956a}
A.~G. Chynoweth, Dynamic method for measuring the pyroelectric effect with
  special reference to barium titanate, Journal of Applied Physics 27~(1)
  (1956) 78--84.
\newblock \href {http://dx.doi.org/10.1063/1.1722201}
  {\path{doi:10.1063/1.1722201}}.

\bibitem{Wawra2023}
J.~Wawra, K.~Nielsch, R.~Hühne, Influence of lattice mismatch on structural
  and functional properties of epitaxial {Ba$_{0.7}$Sr$_{0.3}$TiO$_{3}$} thin
  films, Materials 16 (2023) 6036.
\newblock \href {http://dx.doi.org/10.3390/ma16176036}
  {\path{doi:10.3390/ma16176036}}.

\bibitem{Bowen2014}
C.~Bowen, J.~Taylor, E.~LeBoulbar, D.~Zabek, A.~Chauhan, R.~Vaish, Pyroelectric
  materials and devvicesfor energy harvesting applications, Energy \&
  Enviromental Science 7 (2014) 3836--3856.
\newblock \href {http://dx.doi.org/10.1039/c4ee01759e}
  {\path{doi:10.1039/c4ee01759e}}.

\bibitem{Tong2014}
T.~Tong, J.~Karthik, L.~W. Martin, D.~G. Cahill, Secondary effects in wide
  frequency range measurements of the pyroelectric coefficient of
  {Ba$_{0.6}$Sr$_{0.4}$TiO$_3$} and {PbZr$_{0.2}$Ti$_{0.8}$O$_3$} epitaxial
  layers, Physical Review B 90 (2014) 155423.
\newblock \href {http://dx.doi.org/10.1103/PhysRevB.90.155423}
  {\path{doi:10.1103/PhysRevB.90.155423}}.

\bibitem{Acosta2017}
M.~Acosta, N.~Novak, V.~Rojas, S.~Patel, R.~Vaish, J.~Koruza, G.~Rossetti,
  J.~Rödel, {BaTiO$_3$}-based piezoelectrics: Fundamentals, current status,
  and perspectives, Applied Physics Reviews 4 (2017) 041305.
\newblock \href {http://dx.doi.org/10.1063/1.4990046}
  {\path{doi:10.1063/1.4990046}}.

\bibitem{Vendik2000}
O.~Vendik, S.~Zubko, Ferroelectric phase transition and maximum dielectric
  permittivity of displacement type ferroelectrics {(Ba$_x$Sr$_1-x$TiO$_3$)},
  Jounal of Applied Physics 88 (2000) 5343--5350.
\newblock \href {http://dx.doi.org/10.1063/1.1317243}
  {\path{doi:10.1063/1.1317243}}.

\bibitem{Schl2014a}
D.~G. Schlom, L.-Q. Chen, C.~J. Fennie, V.~Gopalan, D.~A. Muller, X.~Pan,
  R.~Ramesh, R.~Uecker, Elastic strain engineering of ferroic oxides, MRS
  Bulletin 39~(2) (2014) 118–130.
\newblock \href {http://dx.doi.org/10.1557/mrs.2014.1}
  {\path{doi:10.1557/mrs.2014.1}}.

\bibitem{Schm2017a}
M.~Schmidbauer, D.~Braun, T.~Markurt, M.~Hanke, J.~Schwarzkopf, Strain
  engineering of monoclinic domains in {K$_x$Na$_{1-x}$NbO$_3$} epitaxial
  layers: a pathway to enhanced piezoelectric properties, Nanotechnology
  28~(24) (2017) 24LT02.
\newblock \href {http://dx.doi.org/10.1088/1361-6528/aa715a}
  {\path{doi:10.1088/1361-6528/aa715a}}.

\bibitem{schw2017a}
J.~Schwarzkopf, D.~Braun, M.~Hanke, R.~Uecker, M.~Schmidbauer, Strain
  engineering of ferroelectric domains in {K$_x$Na$_{1-x}$NbO$_3$} epitaxial
  layers, Frontiers in Materials 4.
\newblock \href {http://dx.doi.org/10.3389/fmats.2017.00026}
  {\path{doi:10.3389/fmats.2017.00026}}.

\bibitem{Ban2002}
Z.-G. Ban, S.~Alpay, Phase diagrams and dielectric response of epitaxial barium
  strontium titanate films: A theoretical analysis, Jounal of Applied Physics
  91 (2002) 9288--9296.
\newblock \href {http://dx.doi.org/10.1063/1.1473675}
  {\path{doi:10.1063/1.1473675}}.

\bibitem{Liu2016}
Y.~Liu, J.~F. Scott, B.~Dkhil, Direct and indirect measurements on
  electrocaloric effect: Recent developments and perspectives, Applied Physics
  Reviews 3 (2016) 031102.
\newblock \href {http://dx.doi.org/10.1063/1.4958327}
  {\path{doi:10.1063/1.4958327}}.

\bibitem{kutnjak2015i}
Z.~Kutnjak, B.~Rožič, R.~Pirc, Electrocaloric Effect: Theory, Measurements,
  and Applications, John Wiley \& Sons, Ltd, 2015, pp. 1--19.
\newblock \href {http://dx.doi.org/10.1002/047134608X.W8244}
  {\path{doi:10.1002/047134608X.W8244}}.

\bibitem{strukov2003}
B.~Strukov, S.~Davitadze, S.~Kravchun, S.~Taraskin, M.~Goltzman, V.~Lemanov,
  S.~Shulman, Specific heat and heat conductivity of {BaTiO$_3$}polycrystalline
  films in the thickness range 20--1100 nm, Journal of Physics: Condensed
  Matter 15~(25) (2003) 4331.
\newblock \href {http://dx.doi.org/10.1088/0953-8984/15/25/304}
  {\path{doi:10.1088/0953-8984/15/25/304}}.

\bibitem{Kriegner2025}
D.~Kriegner, E.~Wintersberger,
  \href{https://pypi.org/project/xrayutilities/}{xrayutilities: A python
  package for x-ray diffraction data analysis and simulation}, available at
  PyPI and GitHub: https://github.com/dkriegner/xrayutilities (2025).
\newline\urlprefix\url{https://pypi.org/project/xrayutilities/}

\bibitem{schi2013a}
D.~Schick, R.~Shayduk, A.~Bojahr, M.~Herzog, C.~von Korff~Schmising, P.~Gaal,
  M.~Bargheer, {Ultrafast reciprocal-space mapping with a convergent beam},
  Journal of Applied Crystallography 46~(5) (2013) 1372--1377.
\newblock \href {http://dx.doi.org/10.1107/S0021889813020013}
  {\path{doi:10.1107/S0021889813020013}}.

\bibitem{Schi2021a}
D.~Schick, udkm1dsim – a python toolbox for simulating 1d ultrafast dynamics
  in condensed matter, Computer Physics Communications 266 (2021) 108031.
\newblock \href {http://dx.doi.org/10.1016/j.cpc.2021.108031}
  {\path{doi:10.1016/j.cpc.2021.108031}}.

\bibitem{Kost1998a}
P.~Kostic, Y.~Okada, N.~C. Collins, Z.~Schlesinger, J.~W. Reiner, L.~Klein,
  A.~Kapitulnik, T.~H. Geballe, M.~R. Beasley, Non-fermi-liquid behavior of
  $\mathrm{SrRuO}{}_{3}$: Evidence from infrared conductivity, Phys. Rev. Lett.
  81 (1998) 2498--2501.
\newblock \href {http://dx.doi.org/10.1103/PhysRevLett.81.2498}
  {\path{doi:10.1103/PhysRevLett.81.2498}}.

\bibitem{Gaal2023a}
P.~Gaal, D.~Schmidt, M.~Khosla, C.~Richter, P.~Boesecke, D.~Novikov,
  M.~Schmidbauer, J.~Schwarzkopf,
  \href{https://doi.org/10.1016/j.apsusc.2022.155891}{Self-stabilization of the
  equilibrium state in ferroelectric thin films}, Applied Surface Science 613
  (2023) 155891.
\newblock \href {http://dx.doi.org/10.1016/j.apsusc.2022.155891}
  {\path{doi:10.1016/j.apsusc.2022.155891}}.
\newline\urlprefix\url{https://doi.org/10.1016/j.apsusc.2022.155891}

\bibitem{warr1990a}
B.~E. Warren, X-Ray Diffraction, Dover Publications, New York, 1990.

\bibitem{Matt2021a}
M.~Mattern, J.-E. Pudell, G.~Laskin, A.~von Reppert, M.~Bargheer, Analysis of
  the temperature- and fluence-dependent magnetic stress in laser-excited
  {SrRuO$_3$}, Structural Dynamics 8~(2) (2021) 024302.
\newblock \href {http://dx.doi.org/10.1063/4.0000072}
  {\path{doi:10.1063/4.0000072}}.

\bibitem{Bohj2012a}
A.~Bojahr, D.~Schick, L.~Maerten, M.~Herzog, I.~Vrejoiu, C.~von
  Korff~Schmising, C.~Milne, S.~L. Johnson, M.~Bargheer, Comparing the
  oscillation phase in optical pump-probe spectra to ultrafast x-ray
  diffraction in the metal-dielectric {SrRuO${}_{3}$/SrTiO${}_{3}$}
  superlattice, Phys. Rev. B 85 (2012) 224302.
\newblock \href {http://dx.doi.org/10.1103/PhysRevB.85.224302}
  {\path{doi:10.1103/PhysRevB.85.224302}}.

\bibitem{Hubm2016a}
A.~H. Hubmann, S.~Li, S.~Zhukov, H.~von Seggern, A.~Klein, Polarisation
  dependence of schottky barrier heights at ferroelectric {BaTiO$_3$/RuO$_2$}
  interfaces: influence of substrate orientation and quality, Journal of
  Physics D: Applied Physics 49~(29) (2016) 295304.

\bibitem{gros2014a}
R.~Gross, A.~Marx, Festk{\"o}rperphysik, De Gruyter, Berlin, 2014.

\bibitem{Hart2000a}
A.~J. Hartmann, M.~Neilson, R.~N. Lamb, K.~Watanabe, J.~F. Scott, Ruthenium
  oxide and strontium ruthenate electrodes for ferroelectric thin-films
  capacitors, Applied Physics A 70~(2) (2000) 239--242.
\newblock \href {http://dx.doi.org/10.1007/s003390050041}
  {\path{doi:10.1007/s003390050041}}.

\bibitem{Free2016a}
C.~R. Freeze, S.~Stemmer, Role of film stoichiometry and interface quality in
  the performance of {(Ba,Sr)TiO$_3$} tunable capacitors with high figures of
  merit, Applied Physics Letters 109~(19) (2016) 192904.
\newblock \href {http://dx.doi.org/10.1063/1.4967374}
  {\path{doi:10.1063/1.4967374}}.

\bibitem{Lee2016a}
C.-H. Lee, Y.-J. Oh, D.~Y. Lee, D.-J. Choi, Influence of annealing temperature
  on the dielectric properties of {(Ba,Sr)TiO$_{3}$} thin films deposited on
  various substrates, Journal of the Korean Physical Society 69~(10) (2016)
  1571--1574.
\newblock \href {http://dx.doi.org/10.3938/jkps.69.1571}
  {\path{doi:10.3938/jkps.69.1571}}.

\bibitem{Xi2017a}
Z.~Xi, J.~Ruan, C.~Li, C.~Zheng, Z.~Wen, J.~Dai, A.~Li, D.~Wu, Giant tunnelling
  electroresistance in metal/ferroelectric/semiconductor tunnel junctions by
  engineering the schottky barrier, Nature Communications 8~(1) (2017) 15217.
\newblock \href {http://dx.doi.org/10.1038/ncomms15217}
  {\path{doi:10.1038/ncomms15217}}.

\bibitem{wei2017a}
T.-C. Wei, H.-P. Wang, H.-J. Liu, D.-S. Tsai, J.-J. Ke, C.-L. Wu, Y.-P. Yin,
  Q.~Zhan, G.-R. Lin, Y.-H. Chu, J.-H. He, Photostriction of strontium
  ruthenate, Nature Communications 8~(1) (2017) 15018.
\newblock \href {http://dx.doi.org/10.1038/ncomms15108}
  {\path{doi:10.1038/ncomms15108}}.

\bibitem{Guo2021a}
J.~Guo, W.~Chen, H.~Chen, Y.~Zhao, F.~Dong, W.~Liu, Y.~Zhang, Recent progress
  in optical control of ferroelectric polarization, Advanced Optical Materials
  9~(23) (2021) 2002146.
\newblock \href {http://dx.doi.org/https://doi.org/10.1002/adom.202002146}
  {\path{doi:https://doi.org/10.1002/adom.202002146}}.

\bibitem{Plec2024a}
A.~Plech, P.~Gaal, D.~Schmidt, M.~Levantino, M.~Daniel, S.~Stankov, G.~Buth,
  M.~Albrecht, Laser-initiated electron and heat transport in gold-skutterudite
  {CoSb$_3$} bilayers resolved by pulsed x-ray scattering, New Journal of
  Physics 26~(10) (2024) 103024.
\newblock \href {http://dx.doi.org/10.1088/1367-2630/ad8674}
  {\path{doi:10.1088/1367-2630/ad8674}}.

\bibitem{Herz2022a}
M.~Herzog, A.~von Reppert, J.-E. Pudell, C.~Henkel, M.~Kronseder, C.~H. Back,
  A.~A. Maznev, M.~Bargheer, Phonon-dominated energy transport in purely
  metallic heterostructures, Advanced Functional Materials 32~(41) (2022)
  2206179.
\newblock \href {http://dx.doi.org/10.1002/adfm.202206179}
  {\path{doi:10.1002/adfm.202206179}}.

\bibitem{Lang1974a}
D.~V. Lang, Deep‐level transient spectroscopy: A new method to characterize
  traps in semiconductors, Journal of Applied Physics 45~(7) (1974) 3023--3032.
\newblock \href {http://dx.doi.org/10.1063/1.1663719}
  {\path{doi:10.1063/1.1663719}}.

\bibitem{Stol1985a}
L.~Stolt, K.~Bohlin, Deep-level transient spectroscopy measurements using high
  schottky barriers, Solid-State Electronics 28~(12) (1985) 1215--1221.
\newblock \href {http://dx.doi.org/10.1016/0038-1101(85)90045-0}
  {\path{doi:10.1016/0038-1101(85)90045-0}}.

\end{thebibliography}
\bibliographystyle{elsarticle-num} 
\end{document}